\documentclass[seceq,preprint]{ptptex}

\usepackage{graphicx}
\preprintnumber[3cm]{%<-- [..]: optional width of preprint # column.
arXiv:0902.0445\\
RIKEN-TH-148\\ February, 2009}
\title{%        %You can use \\ for explicit line-break.
Numerical Evaluation of Gauge Invariants for \\
$a$-Gauge Solutions in Open String Field Theory%
}

\author{%       %Use \scshape for the family name.
Isao \textsc{Kishimoto}$^1$ and Tomohiko \textsc{Takahashi}$^2$%
}

\inst{%     %Affiliation, neglected when [addenda] or [errata].
$^1$Theoretical Physics Laboratory, RIKEN,
Wako 351-0198, Japan\\
$^2$Department of Physics, Nara Women's University,
Nara 630-8506, Japan
}

\abst{%         %This abstract is neglected when [addenda] or [errata].
We evaluate gauge invariants (vacuum energy and gauge invariant
overlap) for numerical classical solutions in the cubic open string
field theory under Asano and Kato's $a$-gauge fixing condition. 
We propose an
efficient iterative procedure for solving the equations of motion so
that the BRST invariance of the solutions is numerically ensured. 
The resulting gauge invariants are numerically stable and 
almost equal to
those of Schnabl's tachyon vacuum solution
in the well-defined region of a gauge parameter. 
These results provide further evidence that the
numerical and analytical solutions are gauge-equivalent.
}

\begin{document}

\maketitle

\section{Introduction}

The cubic bosonic open string field theory (SFT)\cite{Witten:1985cc} has
classical solutions 
that are expected to be a tachyon vacuum solution conjectured by
Sen \cite{Sen:1999mh,Sen:1999mg,Sen:1999xm}. One of them is given as a
numerical solution using a level truncation 
scheme in the Siegel gauge \cite{Sen:1999nx,Moeller:2000xv,Gaiotto:2002wy}
and then an analytic solution is constructed in 
the Schnabl gauge,\cite{Schnabl:2005gv} which is a modified version of
the Siegel gauge. These two solutions are believed to be the same
tachyon vacuum 
solution. Although it is difficult to prove their equivalence by
constructing an explicit gauge transformation between them, we can provide
evidence using three gauge invariant quantities. First, we can find
that these two solutions have almost the same 
values for two gauge invariants: one of these invariants is a vacuum 
energy\cite{Schnabl:2005gv} that should precisely  cancel the D-brane
tension and 
the other is a gauge invariant
overlap\cite{Kawano:2008ry,Ellwood:2008jh} that
corresponds to the coupling of an open string field to an
on-shell closed string.
The third gauge invariant is an on-shell scattering amplitude
in the SFT expanded around the tachyon vacuum.
It should exactly vanish since the analytic
solution gives a trivial cohomology of the kinetic operator around the
vacuum\cite{Ellwood:2006ba}. The numerical solution has a similar
tendency to provide vanishing scattering
amplitudes\cite{Imbimbo:2006tz,Giusto:2003wc}. These are results
consistent with the expectation that the analytic and numerical
solutions are equivalent up to gauge transformation.  

The purpose of this study is to provide further evidence of the gauge
equivalence of the two solutions by the numerical calculation of the vacuum
energy and gauge invariant overlap. In our
calculation, we will truncate the level of the string
field and fix it in Asano and Kato's $a$-gauge\cite{Asano:2008iu,Asano:2006hk}. 

The ``$a$-gauge'' was proposed as a family of gauges with one-parameter $a$,
which corresponds to the covariant gauge in the conventional gauge theory.
It includes the Feynman-Siegel gauge ($a=0$) and Landau gauge ($a=\infty$).
In SFT under $a$-gauge fixing condition, it was proved that on-shell
physical amplitudes are gauge-independent\cite{Asano:2008iu}. 
The $a$-gauge condition is also applicable to the numerical analysis of the
tachyon vacuum using the level truncation scheme\cite{Asano:2006hm}.
The potential in the $a$-gauge
has a nontrivial local minimum where
the energy density approximately equals that of the tachyon vacuum.
In other words, the nontrivial vacuum energy remains almost the same as
that of the Siegel gauge for various values of $a$.
Although the $a$-gauge yields good results in level truncation
analysis, the BRST invariance of the vacuum\cite{Hata:2000bj} has not
yet been evaluated and it must be checked to confirm that the vacuum is
truly physical. In this study, we will perform a numerical test on the
BRST invariance, 
or the validity of the classical equations of motion, for the nontrivial
vacuum in the $a$-gauge.

The gauge invariant overlap is an interesting quantity since it
takes nontrivial values for the nonperturbative
vacuum\cite{Kawano:2008ry,Ellwood:2008jh}. Moreover,
from the intensive study of the overlap, we have new insights into the
relation among the tachyon vacuum solution and boundary
states\cite{Kawano:2008jv,Kishimoto:2008zj,Kiermaier:2008qu}.  
In this paper, we will numerically calculate the gauge invariant overlap
for the nontrivial solution in the $a$-gauge, and we will confirm that
it is also in good agreement with the Siegel gauge result.
This will also confirm the gauge equivalence between the numerical and
analytical solutions.

First, we will discuss the string field theory and the equations of motion
focusing on the $a$-gauge fixing condition in \S \ref{sec:eom}. 
In \S\ref{sec:iteration}, we will discuss an iterative algorithm solving
the equations of motion. We will propose a new algorithm that
simplifies numerical computations in the $a$-gauge. 
The numerical results will be provided in
\S \ref{sec:truncated} and then we will give a summary and discussion in
\S \ref{sec:concluding}.
In Appendix \ref{sec:norm}, we will define a norm of string fields. In
Appendix \ref{sec:sample}, we will give samples of numerical data
corresponding to plots in \S \ref{sec:truncated}.

\section{Equations of motion in various gauges
\label{sec:eom}}

In cubic bosonic open SFT\cite{Witten:1985cc}, the gauge-invariant
action is given by 
\begin{eqnarray}
 S[\Psi]=-\frac{1}{g^2}\int\left(\frac{1}{2}\Psi*Q_{\rm B}\Psi
+\frac{1}{3}\Psi*\Psi*\Psi\right),
\end{eqnarray}
where the string field $\Psi$ is expanded by string Fock space states
with the ghost number $1$. 
The action is invariant under the infinitesimal gauge transformation
\begin{eqnarray}
 \delta\Psi=Q_{\rm B}\Lambda+\Psi*\Lambda-\Lambda*\Psi.
\end{eqnarray}
From the least-action principle, the equation of motion is derived as
\begin{eqnarray}
\label{Eq:eom}
 Q_{\rm B}\Psi+\Psi*\Psi=0.
\end{eqnarray}

We impose a gauge fixing condition to solve the equations of
motion. Let us consider the linear gauge fixing
condition~\cite{Asano:2008iu} for some operator ${\cal O}_{\rm GF}$,
\begin{eqnarray}
\label{Eq:gfc}
 {\cal O}_{\rm GF}\Psi=0.
\end{eqnarray}
It provides the Siegel gauge if $b_0$ is taken as ${\cal O}_{\rm GF}$.
The $a$-gauge fixing condition\cite{Asano:2006hk} is defined by the 
operator 
\begin{eqnarray}
 {\cal O}_{\rm GF}=b_0 M + a b_0 c_0 \tilde{Q},
\label{eq:a-gauge_cond}
\end{eqnarray}
where $a$ denotes the gauge parameter and the operators $M$ and
$\tilde{Q}$ are defined by the expansion of the BRST charge with
respect to ghost zero modes,
\begin{eqnarray}
 Q_{\rm B}=\tilde{Q}+c_0 L_0+b_0 M.
\end{eqnarray}
The $a$-gauge
at $a=0$ is proved to be equivalent to the Siegel gauge,
though the operator ${\cal O}_{\rm GF}$ is different from $b_0$.
In the infinite $a$ limit,
the $a$-gauge represents the
Landau gauge for a massless vector field. The Landau gauge can be also
given by the regular operator
\begin{eqnarray}
 {\cal O}_{\rm GF}=b_0 c_0 \tilde{Q}.
\label{eq:Landau-gauge_cond}
\end{eqnarray}

Let us consider classical solutions of the equation of motion
(\ref{Eq:eom}) under 
the gauge fixing condition (\ref{Eq:gfc}).
First, we introduce the undetermined multiplier string field
${\cal B}$ with a ghost number $2-{\rm gh}({\cal O}_{\rm
GF})$, where ${\rm gh}(A)$ denotes the ghost number of $A$, into
the action: 
\begin{eqnarray}
 S[\Psi,{\cal B}]=-\frac{1}{g^2}\int\left(\frac{1}{2}\Psi*Q_{\rm B}\Psi
+\frac{1}{3}\Psi*\Psi*\Psi\right)+\int {\cal B}*{\cal O}_{\rm GF}\Psi.
\label{eq:action+Bpsi}
\end{eqnarray}
The equations of motion are derived as
\begin{eqnarray}
&& {\cal O}_{\rm GF}\Psi=0,
\label{eq:gauge_cond}
\\
&& Q_{\rm B}\Psi+\Psi*\Psi=g^2\,{\rm bpz}({\cal O}_{\rm GF}){\cal B},
\label{eq:EOM=B}
\end{eqnarray}
where ${\rm bpz}({\cal O})$ denotes the BPZ conjugation of some operator
${\cal O}$.

If we find a projection operator ${\cal P}_{\rm GF}$ corresponding to
 the gauge condition (\ref{eq:gauge_cond}), such as
\begin{eqnarray} 
&&{\cal P}_{\rm GF}^2|F\rangle_1={\cal P}_{\rm GF}|F\rangle_1,
~~~~{\cal O}_{\rm GF}{\cal P}_{\rm GF}|F\rangle_1=0,
\label{eq:Projection}
\end{eqnarray}
for any state $|F\rangle_1$ with ${\rm gh}(|F\rangle_1)=1$, we obtain
\begin{eqnarray}
\label{Eq:Peom}
 {\rm bpz}({\cal P}_{\rm GF})\left(
Q_{\rm B}\Psi+\Psi*\Psi\right)=0
\end{eqnarray}
{}from Eq.~(\ref{eq:EOM=B}).
This equation
represents part of the gauge unfixed equation of motion
(\ref{Eq:eom}).\footnote{
Equations (\ref{eq:gauge_cond}) and (\ref{Eq:Peom}) correspond to 
the equations of motion for the gauge fixed action: $S[\Psi]|_{{\cal O}_{\rm
GF}\Psi=0}$, which is obtained by integrating out ${\cal
B}$ in Eq.~(\ref{eq:action+Bpsi}).
}
For the Siegel gauge condition, the operator ${\cal P}_{\rm GF}$
is given by $b_0c_0$.

The projection ${\cal P}_{\rm GF}$ for 
the $a$-gauge Eq.~(\ref{eq:a-gauge_cond}) is given by\cite{Asano:2006hk}
\begin{eqnarray}
{\cal P}_{\rm GF}=1+\frac{1}{a-1}
\Bigl(\frac{\tilde Q}{L_0}+c_0\Bigr)(b_0+a b_0c_0W_1\tilde Q),
\label{Eq:Pgf}
\end{eqnarray}
where we consider $L_0\ne 0$ sector and $a\ne 1$.\footnote{In the case
$a=1$, 
${\cal O}_{\rm GF}=b_0c_0 Q_{\rm B}$ 
is ill-defined as a gauge fixing condition at the free
level.\cite{Asano:2006hk}}
The operator $W_1$ is given by
\begin{eqnarray}
&&W_1=\sum_{i=0}^\infty \frac{(-1)^i}{\{(i+1)!\}^2}
  M^i(M^-)^{i+1},~~
 M^-=-\sum_{n=1}^\infty\frac{1}{2n}b_{-n}b_n.
\end{eqnarray}
Using $\tilde Q^2=-L_0M$ and the relations\cite{Asano:2006hk,Asano:2008iu}
\begin{eqnarray}
&&{\rm bpz}(W_1)b_0c_0|F\rangle_2=-W_1b_0c_0|F\rangle_2,~~~~
{\rm bpz}(W_1)c_0b_0|F\rangle_3=-W_1c_0b_0|F\rangle_3\,,\\
&&MW_1b_0c_0|F\rangle_2=b_0c_0|F\rangle_2,~~~~~~~~~~
MW_1c_0b_0|F\rangle_3=c_0b_0|F\rangle_3,\\
&&W_1Mb_0c_0|F\rangle_0=b_0c_0|F\rangle_0,~~~~~~~~~~
W_1Mc_0b_0|F\rangle_1=c_0b_0|F\rangle_1,
\end{eqnarray}
for any state $|F\rangle_n$ with ${\rm gh}(|F\rangle_n)=n$,
we can derive Eq.~(\ref{eq:Projection}).

\section{Iterative procedure in various gauges
\label{sec:iteration}}

Gaiotto and Rastelli\cite{Gaiotto:2002wy} pointed out that, in the
Siegel gauge, an efficient numerical approach to solving the equations of
motion is Newton's method. As they noted, the iterative algorithm can be 
expressed in the compact form
\begin{eqnarray}
\label{Eq:Newtonmethod}
 \Psi_{(n+1)}=Q_{\Psi_{(n)}}^{-1}(\Psi_{(n)}*\Psi_{(n)})
\end{eqnarray}
for $n=0,1,2,\cdots$ with an initial value $\Psi_{(0)}$.
Here, the operator $Q_{\Psi}^{-1}$ is defined by
\begin{eqnarray}
\label{Eq:SiegelEq1}
&& b_0Q_{\Psi}^{-1}|F\rangle_2=0,\\
\label{Eq:SiegelEq2}
&&c_0b_0\left(Q_{\Psi}Q_{\Psi}^{-1}|F\rangle_2-|F\rangle_2\right)=0
\end{eqnarray}
for any ghost number two state $|F\rangle_2$, where
\begin{eqnarray}
 Q_{\Psi}\Phi=Q_{\rm
  B}\Phi+\Psi*\Phi+\Phi*\Psi
\end{eqnarray}
for any ghost number one string field $\Phi$.
Eq.~(\ref{Eq:SiegelEq1}) corresponds to the Siegel gauge condition and 
Eq.~(\ref{Eq:SiegelEq2}) implies that $Q_{\Psi}^{-1}$ is
an inverse operator of $Q_{\Psi}$ in a sense.

Noting the above definition, if we get an
$n$-th configuration $\Psi_{(n)}$, we can construct an $(n+1)$-th
configuration $\Psi_{(n+1)}$ by solving
the linear equations
\begin{eqnarray}
\label{eq:Siegel_gauge_n+1}
&& b_0 \Psi_{(n+1)}=0,\\
&&
 c_0b_0\left(Q_{\Psi_{(n)}}\Psi_{(n+1)}-\Psi_{(n)}*\Psi_{(n)}\right)=0.
\label{eq:c0b0QPsin+1}
\end{eqnarray}
Suppose that we find a converged configuration $\Psi_{(\infty)}$ after the
infinite iterative process of Eq.~(\ref{Eq:Newtonmethod}). 
By Eqs.~(\ref{eq:Siegel_gauge_n+1}) and (\ref{eq:c0b0QPsin+1}), we can find
 that the obtained $\Psi_{(\infty)}$ satisfies the Siegel gauge condition
$b_0\Psi_{(\infty)}=0$ and
\begin{eqnarray}
&&c_0b_0(Q_{\rm B}\Psi_{(\infty)}+\Psi_{(\infty)}*\Psi_{(\infty)})=0.
\label{eq:c0b0eom_Siegel}
\end{eqnarray}
This is a projected part of the whole equation of motion (\ref{Eq:eom}),
which is equivalent to Eq.~(\ref{Eq:Peom})
with ${\cal P}_{\rm GF}=b_0c_0$ in the Siegel gauge.

Now, let us look for the iterative algorithm applied to 
the $a$-gauge condition.
To do that, we have only to generalize the linear equations
(\ref{eq:Siegel_gauge_n+1}) and (\ref{eq:c0b0QPsin+1}) straightforwardly
using ${\cal O}_{\rm GF}$ given by Eq.~(\ref{eq:a-gauge_cond}) 
(or Eq.~(\ref{eq:Landau-gauge_cond}) for $a=\infty$)
and ${\cal P}_{\rm GF}$ given by Eq.~(\ref{Eq:Pgf}):
\begin{eqnarray}
\label{eq:AKEq7}
&& {\cal O}_{\rm GF}\Psi_{(n+1)}=0,\\
\label{eq:AKEq8}
&& {\rm bpz}({\cal P}_{\rm
GF})\left(Q_{\Psi_{(n)}}\Psi_{(n+1)}-\Psi_{(n)}*\Psi_{(n)}\right)=0.
\end{eqnarray}
Actually, we can find a numerically converged solution through this
algorithm with the initial configuration given in
Eq.~(\ref{eq:Psi0}) using the level truncation.
We can also obtain the same configuration by solving the equation of
motion for the gauge fixed action: $S[\Psi]|_{{\cal O}_{\rm GF}\Psi=0}$,
which is equivalent to Eq.~(\ref{Eq:Peom}) with Eq.~(\ref{Eq:Pgf}),
using the level truncation.

However, Eq.~(\ref{eq:AKEq8}) is so complicated that
the computational speed may be lower than that in the Siegel gauge
case, Eq.~(\ref{eq:c0b0QPsin+1}).  Since the operators $\tilde{Q}$ and
$W_1$ in the projection operator include an infinite sum of ghost modes,
it is a very cumbersome procedure to act these on a state.
As an alternative, let us consider the equations
\begin{eqnarray}
\label{eq:AKEq3}
&& {\cal O}_{\rm GF}\Psi_{(n+1)}=0,\\
\label{eq:AKEq4}
&&
c_0b_0\left(Q_{\Psi_{(n)}}\Psi_{(n+1)}-\Psi_{(n)}*\Psi_{(n)}\right)=0,
\end{eqnarray}
where the projection operator ${\cal P}_{\rm GF}$ (\ref{Eq:Pgf})
is replaced with the simple operator $b_0c_0$. From Eqs.~(\ref{eq:AKEq3}) and
(\ref{eq:AKEq4}), 
a converged configuration $\Psi_{(\infty)}$ after infinite iterations
satisfies
\begin{eqnarray}
\label{Eq:AKEq5}
&& {\cal O}_{\rm GF}\Psi_{(\infty)}=0,\\
\label{Eq:AKEq6}
&& c_0b_0\left(Q_{\rm
	  B}\Psi_{(\infty)}+\Psi_{(\infty)}*\Psi_{(\infty)}\right)=0. 
\end{eqnarray}
Eq. (\ref{Eq:AKEq5}) imposes that $\Psi_{(\infty)}$ is in 
the $a$-gauge
subspace, but Eq.~(\ref{Eq:AKEq6}) is the same as the equation
of motion under the Siegel gauge condition (\ref{eq:c0b0eom_Siegel}). 
This mismatch in the gauge fixing condition
(except in the case $a=0$, which is equivalent to
the Siegel gauge) seems to suggest that we cannot find a
solution in the $a$-gauge by solving Eqs.~(\ref{eq:AKEq3}) and
(\ref{eq:AKEq4}).

Generically, the classical solution should satisfy all of the equations
of motion, included in Eq.~(\ref{Eq:eom}). 
In the case of the Siegel gauge, $\Psi_{(\infty)}$ obeys
Eq.~(\ref{eq:c0b0eom_Siegel}), which is a projected part of
Eq.~(\ref{Eq:eom}).  
It can be used for finding the classical solution.
However the unprojected equations should be satisfied by $\Psi_{(\infty)}$ to
ensure that the solution is a true vacuum. Actually, the remaining
equation $b_0c_0\left(Q_{\rm
B}\Psi_{(\infty)}+\Psi_{(\infty)}*\Psi_{(\infty)}\right)=0$
indicates the BRST invariance of the classical solution and it
holds to a high accuracy for the numerical solution in the Siegel
gauge.\cite{Hata:2000bj,Gaiotto:2002wy}

Therefore, $\Psi_{(\infty)}$ in Eq.~(\ref{Eq:AKEq6}) is a possible
candidate for the solution 
in the $a$-gauge because it satisfies 
the $a$-gauge condition and part of all the equations of motion. 
To verify whether it is a true solution, we will check the
remaining part:
\begin{eqnarray}
b_0c_0\left(Q_{\rm
B}\Psi_{(\infty)}+\Psi_{(\infty)}*\Psi_{(\infty)}\right)=0,
\label{eq:b0c0eom}
\end{eqnarray}
apart from projected equation of motion (\ref{Eq:AKEq6}). 
Thus, we are able to find the numerical solution more efficiently by
using the simplified equations Eqs.~(\ref{eq:AKEq3}) and (\ref{eq:AKEq4}).

\section{Level-truncated solutions
\label{sec:truncated}}

When applying the algorithm in the previous section,
namely, Eqs.~(\ref{eq:AKEq3}) and (\ref{eq:AKEq4}), to find a solution, 
we have to specify an initial configuration $\Psi_{(0)}$.
We take it as
\begin{eqnarray}
\label{eq:Psi0}
 \Psi_{(0)}=\frac{64}{81\sqrt{3}}\,c_1\left|0\right>,
\end{eqnarray}
which is the nontrivial solution
for the level $(0,0)$ truncation.
To proceed with the iteration numerically,
we use the level truncation approximation corresponding to the
$(L,2L)$ and $(L,3L)$ truncations.
Namely, we truncate the string field to level $L\equiv L_0+1$
and interaction terms, which appear in the star product,
up to total level $2L$ or $3L$.
To terminate the iteration,
we should specify the accuracy limit of convergence.
We define a ``norm'' of a string field $\|\cdot \|$ as in Appendix
\ref{sec:norm} to measure the accuracy.
We terminate the iterative procedure if the relative error reaches
\begin{eqnarray}
\label{Eq:accuracy}
 \frac{\|\Psi_{(n+1)}-\Psi_{(n)}\|}{\|\Psi_{(n)}\|}<10^{-8}.
\end{eqnarray}
For all levels of $L$ and various values of $a$,
the $n$-th configuration reaches this accuracy limit
after 10 iteration steps or less.
At the same time, we explicitly examine whether the resulting solution
satisfies Eq.~(\ref{Eq:AKEq6}) by
calculating the quantity
\begin{eqnarray}
\label{Eq:accuracyPEOM}
\frac{\|c_0b_0(Q_{\rm
B}\Psi_{(n)}+\Psi_{(n)}*\Psi_{(n)})\|}{\|\Psi_{(n)}\|}.
\end{eqnarray}
We verified that this quantity is smaller than $10^{-8}$ for the
resulting solution which satisfies the accuracy limit
Eq.~(\ref{Eq:accuracy}).

\subsection{Evaluation of gauge invariants}

\begin{figure}[b]
 \begin{center}
\includegraphics[width=12cm]{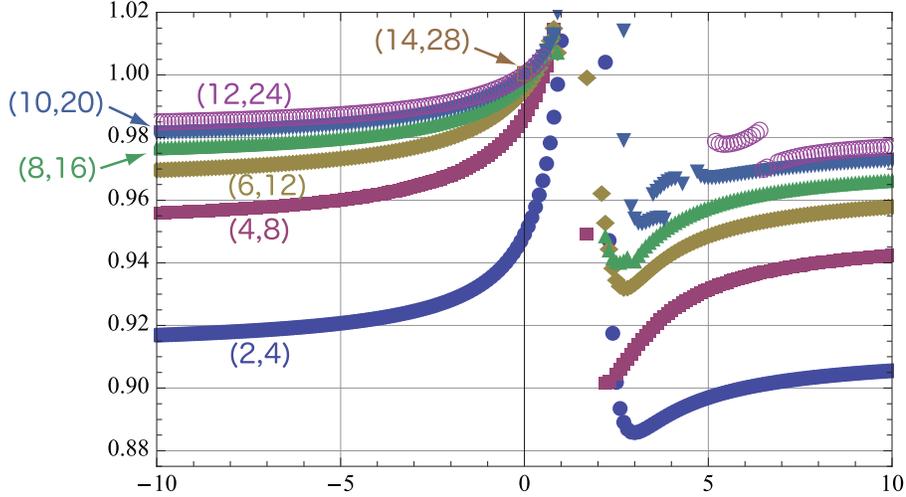}
\end{center}
\caption{Plots of vacuum energy $S(\Psi_a)$
for $(L,2L)$ truncation. The horizontal axis denotes 
the value of the gauge parameter $a$.
(We have only one datum for $(14,28)$ truncation, which is in the Siegel
 gauge $(a=0)$. For other $a$-gauges ($a\ne 0$), calculations are harder
in our {\sl Mathematica} program.)}
\label{fig:veL2L}
\end{figure}

The resulting solution obtained by the iteration above
depends on the gauge parameter. For the solution $\Psi_a$,
we calculate the classical action
\begin{eqnarray}
 S(\Psi_a)=-2\pi^2\left(\frac{1}{2}\left<\Psi_a,\,Q_{\rm B}\Psi_a\right>
+\frac{1}{3}\left<\Psi_a,\,\Psi_a*\Psi_a\right>\right),
\end{eqnarray}
which is normalized to be one for the Schnabl solution.
In Fig.~\ref{fig:veL2L}, we show plots of vacuum energy for $(L,2L)$
truncation as a function of $a$.
\begin{figure}[b]
 \begin{center}
\includegraphics[width=12cm]{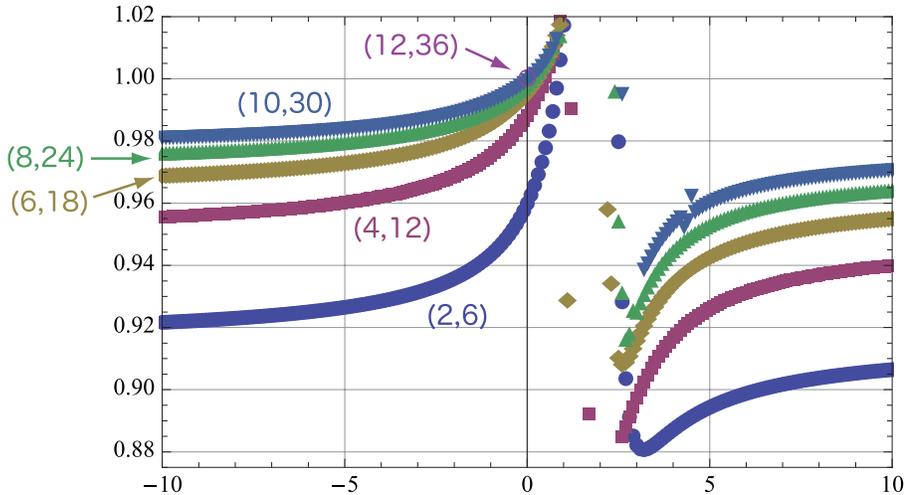}
\end{center}
\caption{Plots of vacuum energy $S(\Psi_a)$
for $(L,3L)$ truncation. The horizontal axis denotes 
the value of the gauge parameter $a$.
(We have only one datum for $(12,36)$ truncation, which is
 in the Siegel gauge $(a=0)$.)}
\label{fig:veL3L}
\end{figure}

In the region at approximately $a=1$, the value of the action is unstable for
every level. This instability was reported to occur for level 2, 4 and 6
analyses in an earlier paper\cite{Asano:2006hm}.
According to the paper, $a=1$ is a gauge
nonfixed point in the free theory and then the nearby gauge horizon 
seems to remain at approximately $a=1$ if the interaction is switched
on. The plots in
Fig.~\ref{fig:veL2L} suggest that the situation would not improve
despite higher-level calculation.

In the well-defined region except the dangerous zone at approximately
$a=1$, the value of the
action is stable at over 90\% of the expected value for the
tachyon vacuum. Moreover, the value gradually approaches $1$ as
truncation level is increased. These are good results, which are consistent
with the gauge independence of vacuum energy. The same tendency
is found in the level $(L,3L)$ calculation, as depicted in
Fig.~\ref{fig:veL3L}. 

\begin{figure}[t]
 \begin{center}
\includegraphics[width=12cm]{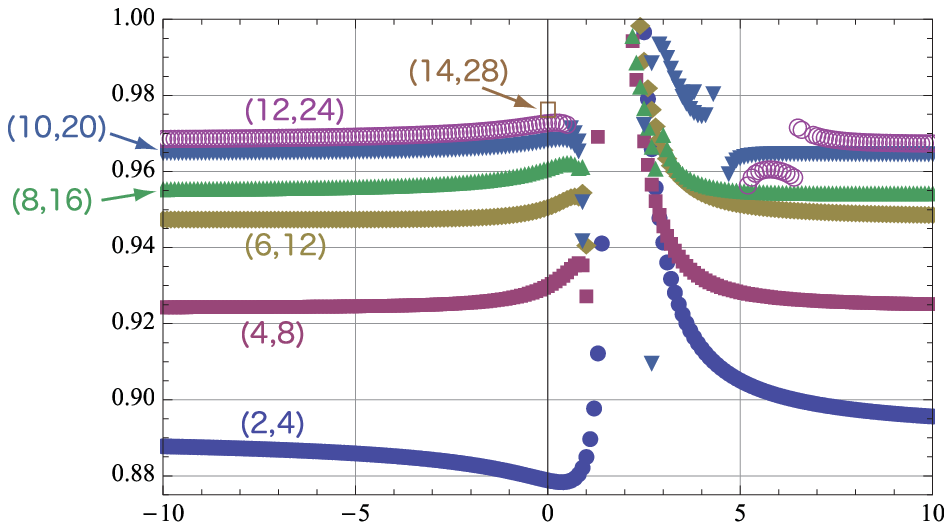}
\end{center}
\caption{Plots of gauge invariant overlap ${\cal O}_V(\Psi_a)$
for $(L,2L)$ truncation. The horizontal axis denotes 
the value of the gauge parameter $a$.
(We have only one datum for $(14,28)$ truncation, which is
 in the Siegel gauge $(a=0)$.)}
\label{fig:overlapL2L}
\end{figure}

\begin{figure}[h]
 \begin{center}
\includegraphics[width=12cm]{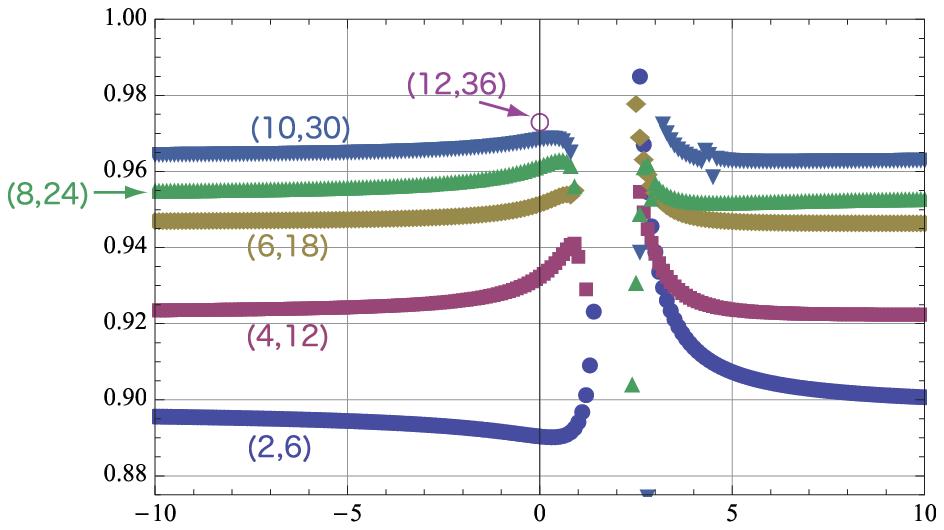}
\end{center}
\caption{Plots of gauge invariant overlap ${\cal O}_V(\Psi_a)$
for $(L,3L)$ truncation. The horizontal axis denotes 
the value of the gauge parameter $a$.
(We have only one datum for $(12,36)$ truncation, 
which is in the Siegel gauge $(a=0)$.)}
\label{fig:overlapL3L}
\end{figure}

Now, let us consider the gauge invariant overlap for the numerical
solution. The gauge invariant overlap is defined by\footnote{See
Ref.~\citen{Kawano:2008ry} for more details.}
\begin{eqnarray}
 {\cal O}_V(\Psi)=\left<{\cal I}\right|
V(i)\left|\Psi\right>,
\end{eqnarray}
where ${\cal I}$ denotes the identity string field, and $V(i)$
corresponds to an on-shell closed string vertex operator.
Hereafter, the overlap is normalized so that it equals $1$ for the
Schnabl tachyon vacuum solution.\footnote{
Namely, we evaluate ${\cal O}_V(\Psi_a)\equiv{\cal
O}_{\eta}(\Psi_a)/{\cal O}_{\eta}(\Psi_{\lambda=1})=2\pi{\cal
O}_{\eta}(\Psi_a)$
with the notation in Ref.~\citen{Kawano:2008ry}.
}
Figs.~\ref{fig:overlapL2L} and \ref{fig:overlapL3L} show
plots of the gauge invariant overlap against $a$ for level $(L,2L)$ and
$(L,3L)$ truncations. As in the case of the action, the plots are
almost gauge-independent in the well-defined region of the gauge
parameter $a$. As 
truncation level is increased, the stable value of the overlap
approaches the expected value of $1$.\footnote{
The approaching speed of the overlap seems to be slower than 
that of the vacuum energy.
}
These results suggest that the
numerical value of the overlap is physically reliable.


\begin{figure}[h]
 \begin{center}
\includegraphics[width=12cm]{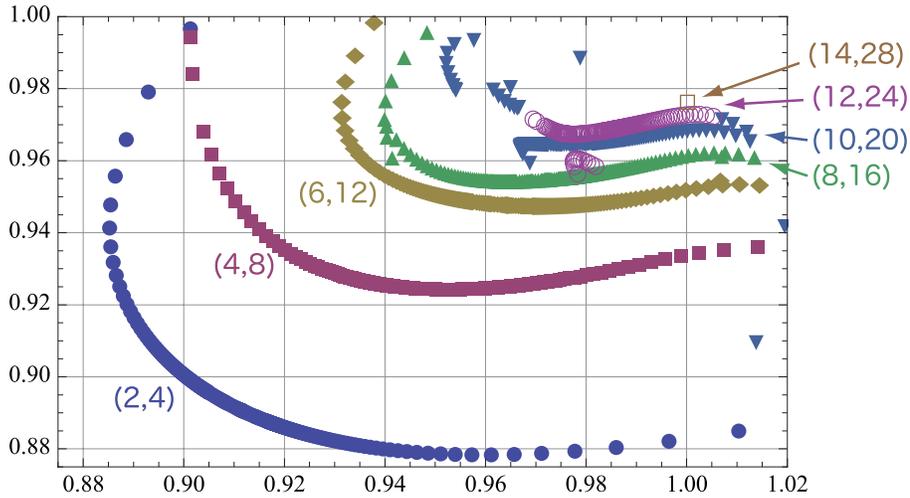}
\end{center}
\caption{Plots of gauge invariants
for $(L,2L)$ truncation. The horizontal axis denotes 
the action $S(\Psi_a)$ and the vertical one denotes the gauge invariant
 overlap ${\cal O}_V(\Psi_a)$. Each point denotes the value of
$(S(\Psi_a),{\cal  O}_V(\Psi_a))$ for various $a$ values.
The left part of the ``curve'' for each level 
corresponds to $4 \lesssim  a < +\infty$ and the right part corresponds to
$-\infty < a\lesssim 1/2$. 
The plots for $a\to +\infty$ and $a\to -\infty$
are continuously connected at that of the Landau gauge ($a=\infty$).
}
\label{fig:ginvL2L}
\end{figure}

\begin{figure}[h]
 \begin{center}
\includegraphics[width=12cm]{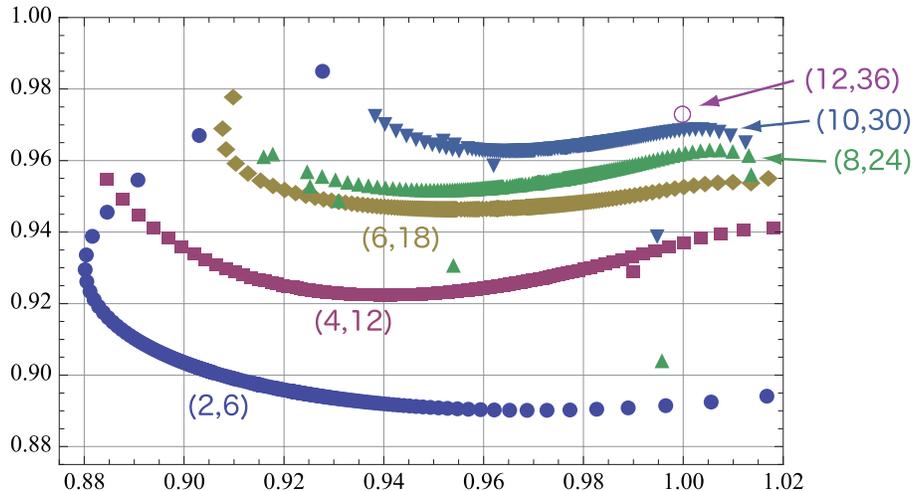}
\end{center}
\caption{Plots of gauge invariants
for $(L,3L)$ truncation. The horizontal axis denotes 
the action $S(\Psi_a)$ and the vertical one denotes the gauge invariant
 overlap ${\cal O}_V(\Psi_a)$. The tendency of the plots is similar to
 that of the $(L,2L)$ truncation in Fig.~\ref{fig:ginvL2L}.}
\label{fig:ginvL3L}
\end{figure}

Here, we display graphs of the action and overlap for various $a$ in
Figs.~\ref{fig:ginvL2L} and \ref{fig:ginvL3L}. The point $(1,1)$
is the result for Schnabl's analytic solution and these figures
clearly indicate that the numerical result from higher-level calculation
is closer to the analytic result for various gauge parameters.

\subsection{The validity of the equation of motion}

\begin{figure}[h]
 \begin{center}
\includegraphics[width=12cm]{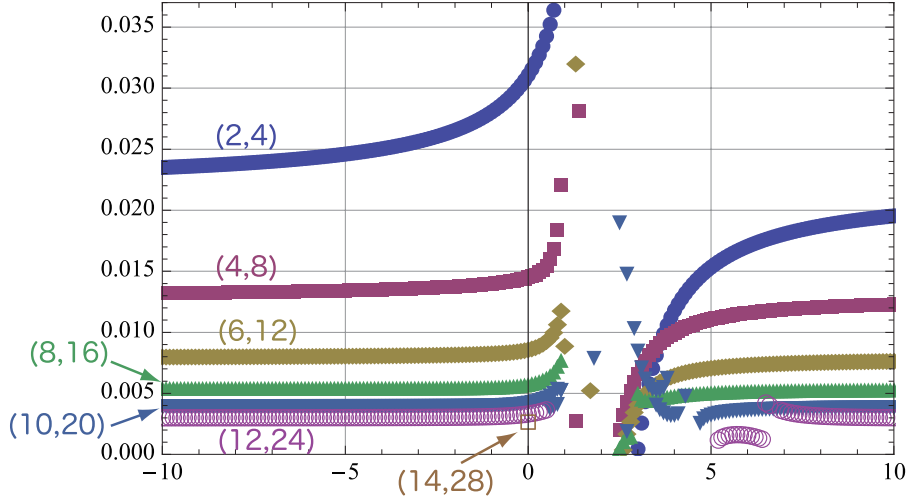}
\end{center}
\caption{Plots of coefficient of $c_{-2}c_1|0\rangle$ 
on the left-hand side of Eq.~(\ref{eq:b0c0eom})
 for $(L,2L)$ truncation.
The horizontal axis denotes $a$.}
\label{fig:cm2eomL2L}
\end{figure}

\begin{figure}[h]
 \begin{center}
\includegraphics[width=12cm]{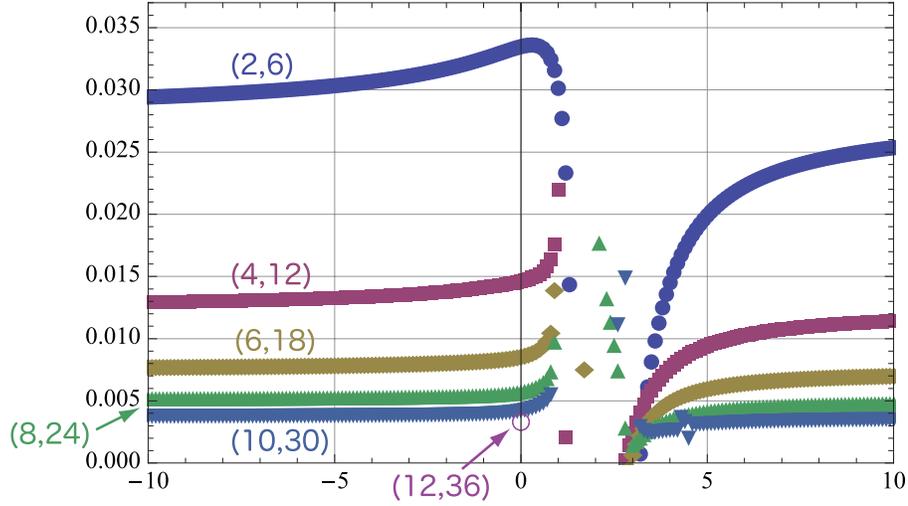}
\end{center}
\caption{Plots of coefficient of $c_{-2}c_1|0\rangle$ 
on the left-hand side of  (\ref{eq:b0c0eom})
for $(L,3L)$ truncation. 
The horizontal axis denotes $a$.}
\label{fig:cm2eomL3L}
\end{figure}

\begin{figure}[h]
 \begin{center}
\includegraphics[width=12cm]{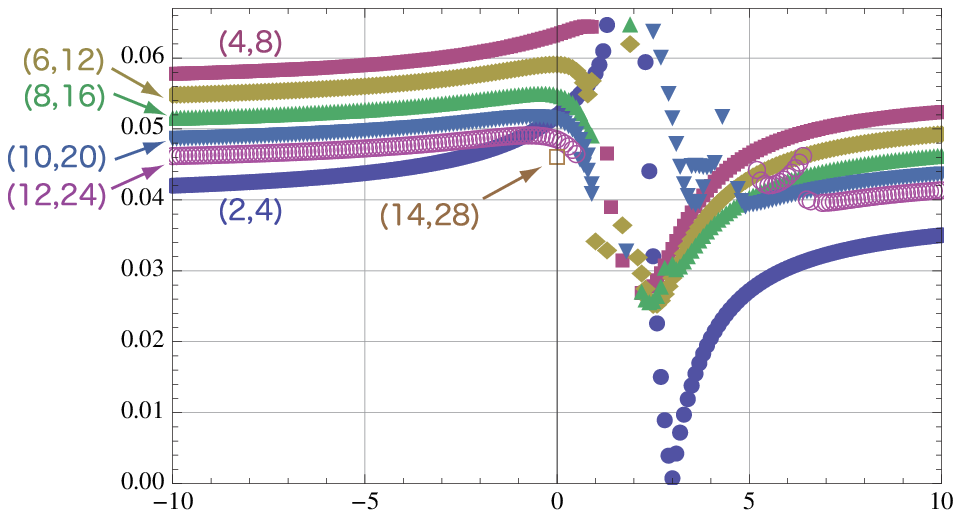}
\end{center}
\caption{Plots of (\ref{Eq:accuracyEOM}) for $(L,2L)$ truncation.
The horizontal axis denotes $a$.
For a fixed $a$, the value of Eq.~(\ref{Eq:accuracyEOM}) decreases 
with increasing level except for the $(2,4)$-truncation.}
\label{fig:eomL2L}
\end{figure}

\begin{figure}[h]
 \begin{center}
\includegraphics[width=12cm]{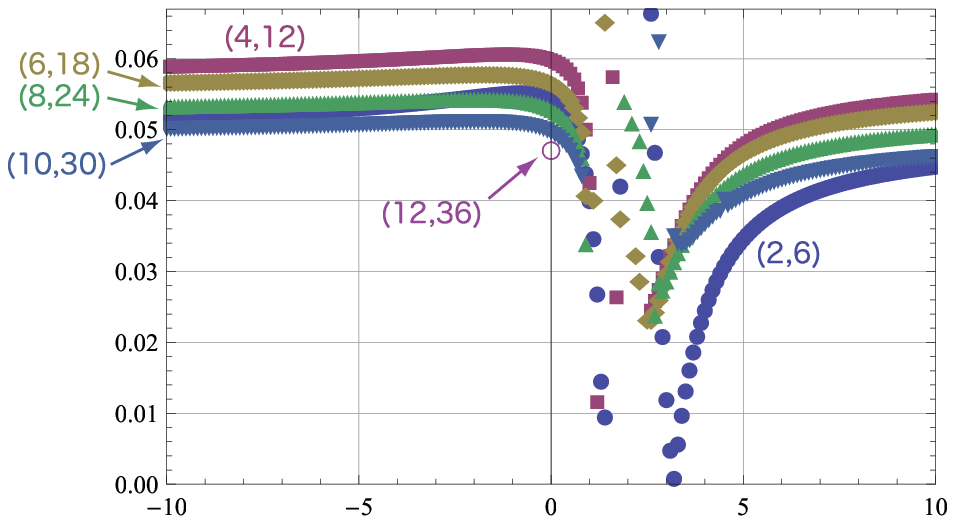}
\end{center}
\caption{Plots of (\ref{Eq:accuracyEOM}) for $(L,3L)$ truncation. 
The horizontal axis denotes $a$.
For a fixed $a$, the value of Eq.~(\ref{Eq:accuracyEOM}) decreases 
with increasing level except for the $(2,6)$-truncation.}
\label{fig:eomL3L}
\end{figure}

Finally, we consider the remaining part of the equations of motion
(\ref{eq:b0c0eom}) for the resulting solution $\Psi_a$.
To check it, let us consider the coefficient of
$c_{-2}c_1|0\rangle$, which is
the lowest-level state included on the 
left-hand side of Eq.~(\ref{eq:b0c0eom}).
We plot it in Fig.~\ref{fig:cm2eomL2L} for the $(L,2L)$ truncation 
and in Fig.~\ref{fig:cm2eomL3L} for the $(L,3L)$ truncation.
Both of them imply that the coefficient approaches zero
at a higher level except in the dangerous zone at approximately $a=1$.
We find that other coefficients on the 
left-hand side of Eq.~(\ref{eq:b0c0eom}) also approach zero
at a higher level.
In order to check all coefficients at one time, we compute
\begin{eqnarray}
\label{Eq:accuracyEOM}
 \frac{\|b_0c_0(Q_{\rm B}\Psi_a+\Psi_a*\Psi_a)\|}{\|\Psi_a\|}.
\end{eqnarray}
This quantity is almost the same as
$\|Q_{\rm B}\Psi_a+\Psi_a*\Psi_a\|/\|\Psi_a\|$
because Eq.~(\ref{Eq:accuracyPEOM}) is negligible as mentioned earlier.
We observe that $\|\Psi_a\|$ is within $0.56\sim 0.7$.
Therefore, Eq.~(\ref{Eq:accuracyEOM}) can be used to measure
the validity of all the equations of motion.
We display the plots of Eq.~(\ref{Eq:accuracyEOM}) for various $a$ values in
Figs.~\ref{fig:eomL2L} and \ref{fig:eomL3L}. Similarly, these plots are
numerically stable in the well-defined region of the gauge parameter. We
find that the norm approaches zero as the level is
increased. 
Thus, the numerical solutions to Eq.~(\ref{Eq:AKEq6}) in the
$a$-gauges constructed using Eqs.~(\ref{eq:AKEq3}), (\ref{eq:AKEq4}) and
(\ref{eq:Psi0}) are the solutions to the equation of motion 
(\ref{Eq:eom}) to a good accuracy.

Here, we should comment on the computational method using the iterative
equations Eqs.~(\ref{eq:AKEq7}) and  (\ref{eq:AKEq8}) with
Eq.~(\ref{eq:Psi0}). Based on these
equations, we can also find numerical solutions for various 
values of the gauge
parameter $a$. The action and overlap for the solutions take
numerical values around those of the analytic result
for Schnabl's solution. However, except that in
the Siegel gauge case ($a=0$), the norm of all the equations of motion
increases for a higher level. 
This suggests that the
resulting solutions become
worse as truncation level increases.
Therefore, we emphasize that the iterative procedure based on
Eqs.~(\ref{eq:AKEq3}) and (\ref{eq:AKEq4}) has a significant advantage
in that the resulting solutions numerically improve the accuracy of
the equation of motion with respect to its norm.


\section{Concluding remarks
\label{sec:concluding}}

We have evaluated gauge invariants (action and gauge invariant overlap)
for numerical solutions 
in the $a$-gauge by level truncation approximation. 
We have checked the validity of the equation of motion
for the solutions.
In the well-defined region of the gauge parameter $a$, the
resulting gauge invariants are numerically equal to those of 
Schnabl's tachyon vacuum  solution.
This provides evidence that previous numerical
results in the Siegel gauge are gauge-independent and thus are
 physically correct. 
The results are consistent with the expectation that
these solutions 
in the $a$-gauge
are gauge-equivalent to Schnabl's
solution and represent a unique nonperturbative vacuum
in bosonic open SFT.

The iterative procedure used in this study to solve the equations of
motion is an efficient algorithm 
in the $a$-gauge.
The algorithm simplifies linear equations 
in the $a$-gauge and achieves a reliable accuracy of
the equation of motion with respect to its norm,
which is nearly equal to that of the Siegel gauge.
It would be interesting to determine why our algorithm is better than 
the conventional calculation method.

In this study, we used a norm with respect to a particular basis in
order to measure the validity of the equation of motion.
However, the norm convergence for a large $L$ limit might be a very
strict condition in the level truncation approximation.
It may be important to investigate the higher-level dependence of
numerical solutions extensively, which will shed
some light on good regularizations of string fields.

\section*{Acknowledgements}

We would like to thank Mitsuhiro Kato for valuable comments.
Discussions during the RIKEN Symposium ``Towards New Developments in
Field and String Theories'' and Sapporo Winter School 2009
 were useful in completing this work.
The work of I.~K. was supported in part by a Special Postdoctoral 
Researchers Program at RIKEN and a Grant-in-Aid for Young Scientists
(\#19740155) from MEXT of Japan.
The work of T.~T. was supported in part by a Grant-in-Aid for 
Young Scientists (\#18740152) from MEXT of Japan.
The level truncation calculations based on {\sl Mathematica} were 
carried out partly on the computer {\it sushiki} at Yukawa Institute 
for Theoretical Physics in Kyoto University.

%\appendix
%\section{First Appendix} %Empty argument \section{} yields `Appendix'. 
%

\appendix

\section{Norm of String Fields
\label{sec:norm}}

Here, we define a norm of string fields to investigate the accuracy
of convergence of the iteration Eq.~(\ref{Eq:accuracy}) and the validity
of the equations of motion, Eqs.~(\ref{Eq:accuracyPEOM}) and
(\ref{Eq:accuracyEOM}),
numerically.
Noting that ${\cal O}_{\rm GF}$ Eq.~(\ref{eq:a-gauge_cond}) (or
Eq.~(\ref{eq:Landau-gauge_cond}) for
$a=\infty$), which specifies the $a$-gauge condition,
is made of the matter Virasoro modes $L_n^{({\rm m})}$ and 
$bc$-ghost modes only and commutes with $L_0$,
we can restrict string fields to twist even universal space
to proceed with the iterations of Eqs.~(\ref{eq:AKEq3}) and
(\ref{eq:AKEq4}) (or (\ref{eq:AKEq8})) with the initial configuration
Eq.~(\ref{eq:Psi0}).

The universal space is spanned by the states whose matter sector is
of the form:
\begin{eqnarray}
 L^{({\rm m})}_{-n_1}L^{({\rm
m})}_{-n_2}\cdots L^{({\rm
m})}_{-n_q}|0\rangle_{\rm m}.~~~~(n_1\ge n_2\ge \cdots \ge n_q\ge 2)
\label{eq:matVirasoro}
\end{eqnarray}
We take an orthonormalized basis with respect to 
the BPZ inner product in the matter sector such as
\begin{eqnarray}
 \langle \varphi_{k,m_k},\varphi_{k',m'_{k'}}\rangle=
(-1)^k\delta_{k,k'}\delta_{m_k,m'_{k'}},~~~~~
L_0^{({\rm
  m})}|\varphi_{k,m_k}\rangle=k|\varphi_{k,m_{k}}
\rangle,
\end{eqnarray}
which is given by appropriate linear combinations of
(\ref{eq:matVirasoro}).
In the ghost sector, we take a basis such as
\begin{eqnarray}
&&|\psi_{k,m_k}\rangle=b_{-p_1}b_{-p_2}\cdots
 b_{-p_r}c_{-q_1}c_{-q_2}\cdots c_{-q_s}c_1|0\rangle_{\rm gh},\\
&&p_1>p_2>\cdots>p_r\ge 1,~~q_1>q_2>\cdots>q_s\ge 0,~~\sum_{t=1}^r p_t
 +\sum_{u=1}^sq_u=k.
\end{eqnarray}
Namely, our basis for twist even universal space is of the form
$\varphi_{k,m_k}\otimes \psi_{l,n_l}$ whose level 
$k+l$ is even.
In the level $(L,2L)$ or $(L,3L)$ truncation, string fields $\Phi$
can be expanded as
\begin{eqnarray}
\Phi=\sum_{k+l\le L}\sum_{m_k,n_l}t_{k,m_k;l,n_l}\,
\varphi_{k,m_k}\otimes \psi_{l,n_l}.
\end{eqnarray}
Using this expansion, we define its norm $\|\Phi\|$ as 
\begin{eqnarray}
 \|\Phi\|=\left(\sum_{k,m_k,l,n_l}|t_{k,m_k;l,n_l}|^2\right)^{\frac{1}{2}}.
\end{eqnarray}

\section{Samples of Numerical Data
\label{sec:sample}}

In the following, we give some data
of our numerical computation with level truncation.
% Table I
\renewcommand{\arraystretch}{.9}
\begin{table}[h]
\caption{Vacuum energy $S(\Psi_a)$
for $(L,2L)$ truncation. 
At $a=4$, the iteration for at least 10 steps
does not converge for $(12,24)$-truncation.
For $(14,28)$ truncation, we computed the configuration for $a=0$ 
only. In the case of the Siegel gauge $(a=0)$,
we reproduced the data in Ref.~\citen{Gaiotto:2002wy} 
up to $L=14$.
}
\label{tab:action_aL2L}
\begin{center}
 \begin{tabular}{|c|c|c|c|c|c|}
\hline
$L$&$a=\infty$&$a=4$&$a=0.5$&$a=0$&$a=-2$\\
\hline
2 & 0.911461&0.891405&0.965684&0.948553&0.927610\\
\hline
4 & 0.949735&0.924272&0.998777&0.986403&0.967567\\
\hline
6 & 0.964287&0.942319&1.00432&0.994773&0.979586\\
\hline
8 & 0.972147&0.951844&1.00541&0.997780&0.985337\\
\hline
10 & 0.977517&0.966292&1.00550&0.999116&0.988741\\
\hline
12&0.981390&--&1.00531&0.999791&0.991016\\
\hline
14&--&--&--&1.00016&--\\
\hline
 \end{tabular}\\
\end{center}
\end{table}
%% Table II
\begin{table}[b]
\caption{Vacuum energy $S(\Psi_a)$
for $(L,3L)$ truncation. 
For the $(12,36)$ truncation, we computed the configuration for $a=0$ 
only. In the case of the Siegel gauge $(a=0)$,
we reproduced the data in Ref.~\citen{Gaiotto:2002wy} 
up to $L=12$.
The data for $a\ne 0$ obtained using Eq.~(\ref{eq:AKEq4})
are not the same as those in Ref.~\citen{Asano:2006hm},
which can be obtained using Eq.~(\ref{eq:AKEq8}).
}
\label{tab:action_aL3L}
\begin{center}
 \begin{tabular}{|c|c|c|c|c|c|}
\hline
$L$&$a=\infty$&$a=4$&$a=0.5$&$a=0$&$a=-2$\\
\hline
2 &0.914683&0.886606&0.977278&0.959377&0.935227\\
\hline
4 &0.948672&0.916240&1.00007&0.987822&0.968273\\
\hline
6 &0.962778&0.933562&1.00434&0.995177&0.979674\\
\hline
8 &0.970986&0.944420&1.00527&0.997930&0.985329\\
\hline
10 &0.976504&0.952494&1.00534&0.999182&0.988719\\
\hline
12&--&--&--&0.999822&--\\
\hline
 \end{tabular}\\
\end{center}
\end{table}
%% Table III
\begin{table}[h]
\caption{Gauge invariant overlap ${\cal O}_V(\Psi_a)$
for $(L,2L)$ truncation. 
The data for $a=0$ (Siegel gauge) were computed in
 Ref.~\citen{Kawano:2008ry} up to $L=10$.
}
\label{tab:ginv_aL2L}
\begin{center}
 \begin{tabular}{|c|c|c|c|c|c|}
\hline
$L$&$a=\infty$&$a=4$&$a=0.5$&$a=0$&$a=-2$\\
\hline
2 & 0.890189&0.912978&0.877969&0.878324&0.882482\\
\hline
4 & 0.923905&0.931557&0.933061&0.929479&0.925121\\
\hline
6 & 0.947283&0.953864&0.952429&0.950175&0.947428\\
\hline
8 & 0.954482&0.956381&0.961994&0.960617&0.957024\\
\hline
10 & 0.964335&0.974321&0.967957&0.967790&0.965723\\
\hline
12&0.967426&--&0.971900&0.972321&0.969986\\
\hline
14&--&--&--&0.976005&--\\
\hline
 \end{tabular}\\
\end{center}
\end{table}
%% Table IV
\begin{table}[h]
\caption{Gauge invariant overlap ${\cal O}_V(\Psi_a)$
for $(L,3L)$ truncation. 
The data for $a=0$ (Siegel gauge) were computed in
 Ref.~\citen{Kawano:2008ry} up to $L=10$.
}
\label{tab:ginv_aL3L}
\begin{center}
 \begin{tabular}{|c|c|c|c|c|c|}
\hline
$L$&$a=\infty$&$a=4$&$a=0.5$&$a=0$&$a=-2$\\
\hline
2 & 0.896934&0.913230&0.889773&0.889862&0.892187\\
\hline
4 & 0.922329&0.925870&0.936626&0.931952&0.925748\\
\hline
6 & 0.946225&0.947629&0.953084&0.951079&0.947946\\
\hline
8 & 0.953680&0.951765&0.962740&0.961175&0.957158\\
\hline
10 &0.963421&0.963255&0.968226&0.968115&0.965796\\
\hline
12&--&--&--&0.972560&--\\
\hline
 \end{tabular}\\
\end{center}
\end{table}
%% Table V
\begin{table}[h]
\caption{Coefficient of $c_{-2}c_1|0\rangle$ on the left-hand side
 of (\ref{eq:b0c0eom}) for $(L,2L)$ truncation.}
\label{tab:cm2_aL2L}
\begin{center}
 \begin{tabular}{|c|c|c|c|c|c|}
\hline
$L$&$a=\infty$&$a=4$&$a=0.5$&$a=0$&$a=-2$\\
\hline
2 &0.0217972&0.0116541&0.0341147&0.0309281&0.0263995\\
\hline
4 &0.0127526&0.00984274&0.0153471&0.0143721&0.0136299\\
\hline
6 &0.00775069&0.00634177&0.00914375&0.00845481&0.00804441\\
\hline
8 &0.00530408&0.00471651&0.00618425&0.00566299&0.00537775\\
\hline
10 &0.00383387&0.00313558&0.00455325&0.00413794&0.00388027\\
\hline
12&0.00287156&--&0.00353302&0.00319962&0.00293373\\
\hline
14&--&--&--&0.00257694&--\\
\hline
 \end{tabular}\\
\end{center}
\end{table}
%% Table VI
\begin{table}[h]
\caption{Coefficient of $c_{-2}c_1|0\rangle$ on the left-hand side
 of (\ref{eq:b0c0eom}) for $(L,3L)$ truncation.}
\label{tab:cm2_aL3L}
\begin{center}
 \begin{tabular}{|c|c|c|c|c|c|}
\hline
$L$&$a=\infty$&$a=4$&$a=0.5$&$a=0$&$a=-2$\\
\hline
2 &0.0278116&0.0143573&0.0333436&0.0333299&0.0315562\\
\hline
4 &0.0122591&0.00754373&0.0150884&0.0145013&0.0136542\\
\hline
6 &0.00733116&0.00476112&0.00903473&0.00841347&0.00791895\\
\hline
8 &0.00497860&0.00341716&0.00618457&0.00564143&0.00528082\\
\hline
10 &0.00360149&0.00256316&0.00457858&0.00412431&0.00380687\\
\hline
12&--&--&--&0.00319231&--\\
\hline
 \end{tabular}\\
\end{center}
\end{table}
%% Table VII
\begin{table}[h]
\caption{(\ref{Eq:accuracyEOM}) for $(L,2L)$ truncation.}
\label{tab:norm_aL2L}
\begin{center}
 \begin{tabular}{|c|c|c|c|c|c|}
\hline
$L$&$a=\infty$&$a=4$&$a=0.5$&$a=0$&$a=-2$\\
\hline
2 &0.0390811&0.0202607&0.0540923&0.0516649&0.0464113\\
\hline
4 &0.0555603&0.0417410&0.0639907&0.0631216&0.0606142\\
\hline
6 &0.0526005&0.0381777&0.0580066&0.0589096&0.0574928\\
\hline
8 &0.0495063&0.0359985&0.0529176&0.0546416&0.0540280\\
\hline
10 &0.0466431&0.0446046&0.0491597&0.0511616&0.0509679\\
\hline
12&0.0441713&--&0.0462595&0.0483385&0.0483964\\
\hline
14&--&--&--&0.0459432&--\\
\hline
 \end{tabular}\\
\end{center}
\end{table}
%% Table VIII
\begin{table}[h]
\caption{(\ref{Eq:accuracyEOM}) for $(L,3L)$ truncation.}
\label{tab:norm_aL3L}
\begin{center}
 \begin{tabular}{|c|c|c|c|c|c|}
\hline
$L$&$a=\infty$&$a=4$&$a=0.5$&$a=0$&$a=-2$\\
\hline
2 &0.0495298&0.0246397&0.0516469&0.0545699&0.0547543\\
\hline
4 &0.0575524&0.0426358&0.0581117&0.0600145&0.0606919\\
\hline
6 &0.0554774&0.0418015&0.0542740&0.0566424&0.0579232\\
\hline
8 &0.0522320&0.0392428&0.0504252&0.0529741&0.0544630\\
\hline
10 &0.0490601&0.0368915&0.0472976&0.0498170&0.0513263\\
\hline
12&--&--&--&0.0471806&--\\
\hline
 \end{tabular}\\
\end{center}
\end{table}

\end{document}